\documentstyle[epsfig,preprint,aps]{revtex}
\begin{document}
\tighten
\title{Model Calculations for the Two--Fragment Electro-Disintegration 
of $^4$He}  
\author{M. Braun, L. L. Howell, S. A. Sofianos} 
\address{University of South Africa, P O Box 392, Pretoria, 0003,
South Africa }
\author{W. Sandhas}
\address{Physikalisches Institut, Unversit\"{a}t Bonn, D-5300 Bonn 1,
Germany}
\sloppy
\date{\today}
\maketitle
\begin{abstract}
Differential cross sections for the electro-disintegration 
process $e + {^4{\rm He}} \longrightarrow {^3{\rm H}}+ p + e'$        
are calculated, using a model in which the final state interaction is
included by means of a nucleon--nucleus (3+1) potential constructed 
via Marchenko inversion. The required bound--state wave functions 
are calculated within the integrodifferential equation approach (IDEA).
In our model the important condition that the initial bound state and
the final scattering state are orthogonal is fulfilled.
The sensitivity of the cross section to the input
$p{^3{\rm H}}$ interaction in certain kinematical regions
is investigated.  The  approach adopted could  be useful in 
reactions involving few cluster systems where effective 
interactions are not well known and exact methods are presently
unavailable. 
Although,  our Plane--Wave Impulse Approximation results
exhibit, similarly to other calculations,  a dip in the five-fold differential 
cross--section around  a missing momentum of $\sim$450\,MeV/$c$,
it is argued that this is an artifact of the omission of re-scattering 
four--nucleon processes. \\

\noindent PACS numbers: 21.45.+v, 25.10.+s, 25.30.Fj

\end{abstract}

\section{Introduction}
In recent years, electro-disintegration processes in light nuclei
have attracted much experimental and theoretical attention.
In the three--nucleon case, early calculations of the electron-proton
coincidence cross section, {\em i.e.}, of the scattered electron detected
in coincidence with the ejected proton, were performed by Griffy and Oakes
using analytical wave functions without inclusion of the final
state interaction (FSI) \cite{Griffy}. Since then, few--body
techniques were used to obtain exact wave functions.
Using the Faddeev formalism, Lehman \cite{Lehman_I} calculated 
the two-- and three--fragment electro-disintegration of $^3$H and
$^3$He for a separable, Yamaguchi-type potential. In the final state the
interactions between the ejected nucleon and the spectator pair were
neglected.

Heimbach {\em et al}.~\cite{Lehman} also studied the two--fragment
electro-disintegration of $^3$He and $^3$H, but with the inclusion of
the FSI. These calculations were based on a method introduced by
Barbour and Phillips \cite{Phillips}, and extended
by Gibson and Lehman \cite{Gibson1}, for the
photo-disintegration of $^3$He. The equations in this approach were
derived by making use of the Alt-Grassberger-Sandhas (AGS) formalism for the
three--nucleon system \cite{AGS}. As in Ref.~\cite{Lehman_I},
a spin-dependent, $s$-wave separable potential of Yamaguchi form was
used in these calculations. In contrast, van  
Meijgaard and Tjon \cite{Tjon} employed the semi-realistic
Malfliet-Tjon (MT~I-III) potential \cite{mt} represented in separable
form via the unitary pole expansion (UPE).

Realistic interactions have also been used to determine the
electro-disintegration cross sections. Schiavilla \cite{OCS}, for
example, employed the Variational Monte Carlo (VMC) method using the
Urbana \cite{Urbana} and Argonne \cite{Wiringa} nucleon--nucleon
(NN) interactions for
determining the bound state wave functions in the
electro-disintegration of $^3$He. His method involved constructing
the final scattering states to be orthogonal to the initial bound state. In a
completely different method, Laget \cite{Laget} employed a
diagrammatic expansion of the scattering amplitude, which can be used
to account for the most important multi-particle effects in the
nuclear medium such as FSI and meson exchange currents (MEC).  
In a more rigorous approach Ishikawa {\em et al}.\ 
\cite{Glockle} calculated the electron--induced two-fragment disintegration
of $^3$He  by using a 34-channel treatment based on Faddeev-type
equations in momentum space. In their work they employed realistic
two--nucleon interactions, namely the Paris \cite{Paris} and Bonn-B
\cite{Bonn} potentials. Later on, Golak {\em at al.}\ \cite{Golak}
extended the work of Ref.~\cite{Glockle} by including higher angular
momentum components of the NN force (up to $j=2$).  

In contrast to the three--nucleon electro-disintegration, calculations for
the four-nucleon  system are computationally much more demanding and
to date no  calculations have been performed to reliably include
scattering states in a rigorous way. One therefore resorts to using 
the approximation 
\begin{equation}
	^{(-)}\langle{\mbox{\boldmath{q}}_\alpha};\Psi_{A-1}| \simeq
	\,^{(-)}\langle
	{\mbox{\boldmath{q}}_\alpha}|\langle \Psi_{A-1}|\,.
\label{eq:scatt}	
\end{equation}
which is often employed in nuclear physics, and has previously been 
used in the  photo-disintegration \cite{Sofianos} and 
electro-disintegration \cite{Schia4} of $^4$He. 
Here $^{(-)}\langle{\mbox{\boldmath{q}}_\alpha};\Psi_{A-1}|$  
denotes the  full scattering state associated with the channel state
$\langle\mbox{\boldmath $q$}_\alpha|\langle\Psi_{A-1}|$ in which the
ejected proton is moving with momentum $\mbox{\boldmath $q$}_\alpha$
relative to the residual nucleus represented by the bound state
$\langle\Psi_{A-1}|$.  In such a model approach extreme care must be
taken to ensure that the approximate scattering states in
(\ref{eq:scatt}) are orthogonal to the initial bound state
$|\Psi_A\rangle$,
\begin{equation}
	^{(-)}\langle{\mbox{\boldmath{q}}_\alpha}|\langle
	\Psi_{A-1}|\Psi_A\rangle = 0\,.
\label{eq:overlap}
\end{equation}
Non-fulfillment of this condition means that the effective scattering 
state, although asymptotically correct,  can be different from the "true" 
wave function in the interior. This in turn means that 
the off shell characteristics of the effective interaction employed
can be completely wrong. This point is crucial 
since in photo- and electro-processes we are concerned with overlap 
integrals that include also the interior part of the wave--function. 
That condition is often simply ignored or some kind of an
orthogonalization method is employed to ensure that it is fulfilled
(see, for example, Ref.~\cite{OCS}). 
The approximation (\ref{eq:scatt}) together with the 
condition (\ref{eq:overlap}) has been used by Schiavilla
\cite{Schia4} within the VMC method (in a fashion similar to the one
employed in $^3$He electro-disintegration) using 
the Argonne two--nucleon and Urbana-VII three-nucleon
\cite{Schiav} interaction models. The optical potential of van Oers
{\em et al.}  \cite{Oers} was employed in this approach to describe the 3+1
interaction. Laget \cite{Laget4} included the FSI and the meson
exchange currents (MEC) using a diagrammatic expansion similar
to the three--nucleon case. Nagorny {\it et al.}\cite{Nagorny}
included the electromagnetic field in the strongly interacting 
system in a relativistic and gauge invariant way, with the FSI being 
incorporated via the pole piece of the p{}$^3$H$\rightarrow$ p{}$^3$H 
scattering t-matrix \cite{Zay}.
 In a more recent work \cite{Leeuwe98} the Schiavilla, the Laget,
 and the Nagorny methods were employed to study  newly obtained
data by van Leeuwe \cite{Koos}. The results thus obtained were interpreted
as an indication  that the dip could be  due to  FSI effects 
and contributions from two--body currents.

The main findings in the aforementioned 
 investigations of three-- and four--nucleon disintegration processes
were that
(a) in certain kinematical regions there is an extreme sensitivity to
the nuclear forces;
(b) there are important differences between the results obtained when
using different reaction models, and (c) in  the four--nucleon case
all model calculations within the Plane--Wave Impulse Approximation
(PWIA) exhibit a dip in the five-fold differential 
cross--section around  a missing momentum of $\sim$450\,MeV which,
being  absent from the experimental data, seems to be an artifact 
of the models.

In this paper, we investigate the above sensitivities of the
cross section to the input nucleon--nucleus interaction in the exclusive
electro-disintegration process
\begin{equation}
	e + {^4{\rm He}} \longrightarrow {^3{\rm H}}+ p + e'\,,       
	\label{eq:reaction2}
\end{equation}
where the  scattered electron and the ejected
nucleon are assumed to be measured in coincidence. 

In our model the interaction between the outgoing proton and the
residual nucleus is determined by using the nucleon-nucleus scattering
data to generate an effective potential via the Marchenko inverse
procedure \cite{AM63,chadsab,Alt}. The potential obtained would be
unique, provided the phase shifts were known at all energies. Since
theoretically the phase shifts can   be calculated reliably only below the
three--body break up threshold while experimentally they are available 
from a limited number of phase shift analyses, 
extrapolations to higher energies are required. 
The most obvious choice appears to be
the one leading to  states which fulfill the orthogonality condition
given by  Eq.~(\ref{eq:overlap}). Thus the dependence of the
interaction on the  extrapolated phase shifts and the resulting effect
on the cross  section is studied. Such an approach 
could  be useful in reactions involving few cluster systems
where the effective interaction  needed to construct 
the scattering wave function, contains a lot of inherent
ambiguities mainly stemming from the limited scattering data.
In contrast, the bound states of the clusters involved can, nowadays,
be calculated quite reliably using a variety of methods (Hyperspherical
Harmonics, Variational {\em etc.}) with  realistic or semirealistic
NN forces. 
In our case the three--body and four--body 
bound--state wave functions are calculated in the integrodifferential
equation approach (IDEA) \cite{IDEA1,Oehm1,Oehm} using 
the MT~I-III potential \cite{mt}. 

This paper is organised as follows: In Sect. \ref{cross} we
describe the cross section for the electro-disintegration 
$^4$He in configuration space. In Sect. \ref{result} the electro-disintegration
of $^4$He is  discussed  and the results of the model described in
this paper are  compared with the available experimental data and
other theoretical  results. Our concluding remarks are presented in  Sect.
\ref{concl}

\section{The Cross Section}
\label{cross}
The electron--proton coincidence cross section is given by
\begin{equation}
  	\frac{d^3\sigma}{dE_f d\Omega_p d\Omega_e} = 
		\frac{\sigma_{\rm M}}{(\hbar
		c)^3 (2\pi)^3}
		\frac{\rho_f}{4E_iE_f\cos^2 \displaystyle{\frac{\theta}{2}}} 
		|{\cal M}(\mbox{\boldmath $q$}_\alpha)|^2
\label{eq:tcross}
\end{equation}
where $\sigma_{\rm M}$ is the Mott differential cross section, 
\begin{equation}
	\sigma_{\rm M} = \frac{e^4\cos^2 
	\displaystyle{\frac{\theta}{2}}}{4 E_i^2 \sin^4
	\displaystyle{\frac{\theta}{2}}}\,.
\end{equation}
$E_i(E_f)$ is the energy of the incoming (outgoing) electron and $\rho_f$
is the relativistic density of states. The transition matrix is given
by 
\begin{equation}
	{\cal M}(\mbox{\boldmath $q$}_\alpha) =
	^{(-)}\langle\mbox{\boldmath
	$q$}_\alpha;\Psi_{A-1}|H|\Psi_A\rangle\,, 
\label{eq:tmatrix}
\end{equation}
where $H$ is the effective Hamiltonian describing the
interaction between an electron and the nucleons. The fragmentation 
considered here is of the  3+1-type while the ejected proton  
moves away with momentum $\mbox{\boldmath $q$}_\alpha$ with respect 
to the residual bound cluster described by the bound state $\Psi_{A-1}$.\\

The  Hamiltonian for the interaction between an electron 
and $A$ nucleons, is that of Mc Voy and van Hove \cite{McVoy} which 
has been previously employed in the electro-disintegration of the 
trinucleon system \cite{Epp,Lehman_I,Lehman}. 
This effective Hamiltonian, correct to order $\hbar^2 q^2/M^2c^2$,
is   
\begin{eqnarray}
	H & = & -\frac{4\pi e^2}{q_\mu^2}\langle u_f|\sum_{j=1}^A
	 \left\{F_{1N}(q_\mu^2)~e^{-i \mbox{\boldmath $q$} \cdot
	 \mbox{\boldmath $x$}_j} \phantom{\frac{q^2}{8M}} 
          \right.\nonumber \\ 
   	&   & - \frac{F_{1N}(q_\mu^2)}{2M} [(\mbox{\boldmath $p$}_j \cdot
   	\mbox{\boldmath 
	 $\alpha$})~ e^{-i \mbox{\boldmath $q$} \cdot \mbox{\boldmath
	 $x$}_j} +  
           e^{-i\mbox{\boldmath $q$} \cdot \mbox{\boldmath $x$}_j}~
	 (\mbox{\boldmath $p$}_j \cdot \mbox{\boldmath $\alpha$})]
	 \nonumber\\ 
	   &   & - i \left[ \frac{F_{1N}(q_\mu^2) + \kappa
	 F_{2N}(q_\mu^2)}{2M}\right]
	 \mbox{\boldmath $\sigma$}_j\cdot (\mbox{\boldmath $q \times
   	\alpha$})~ 
	 e^{-i \mbox{\boldmath $q$} \cdot \mbox{\boldmath $x$}_j} \nonumber\\
	&   & \left.+ \frac{q_\mu^2}{8M^2}~[F_{1N}(q_\mu^2) + 2\kappa
         F_{2N}(q_\mu^2)]~ 
	  e^{-i \mbox{\boldmath $q$} \cdot \mbox{\boldmath $x$}_j} \right \}
	|u_i\rangle\,, 
	\label{eq:hamiltonian}
\end{eqnarray}
where $\mbox{\boldmath $x$}_j$ and
$\mbox{\boldmath $p$}_j$ are vectors denoting the position
and momentum of the $j$-th  nucleon; ${\mbox{\boldmath $\sigma$}}_j$ is
the nucleon spin operator, 
${\mbox{\boldmath $\alpha$}}$ is the electron's Dirac operator acting
on the free electron spinors $|u_i\rangle $ and $|u_f\rangle
$, while $q_\mu^2$ is 
the exchanged four-momentum squared; $F_{1N}$ and $F_{2N}$ are the
Dirac and Pauli form factors of the nucleon, $\kappa$ is the
anomalous moment of the nucleon in nuclear magnetons, and $M$ is the
nucleon mass. For a  proton knock-out the transition matrix,
 Eq.~(\ref{eq:tmatrix}), is written
\begin{eqnarray} 
	{\cal M}
	& = & \langle u_f|\langle f|\,\frac{1}{2}(1+\tau)_j\left[
	F_{1p}(q_\mu^2) + 
	\frac{q_\mu^2}{8M^2}[F_{1p}(q_\mu^2)  + 2\kappa_p F_{2p}(q_\mu^2)]
	\right] e^{-i \mbox{\boldmath $q$} \cdot \mbox{\boldmath $x$}_j}
	|\Psi_A \rangle|u_i\rangle  \nonumber \\
	& & - ~\langle u_f|\langle f|\mbox{\boldmath $\alpha$} \cdot  \left
	[\,\frac{1}{2}(1+\tau)_j ~\frac{F_{1p}}{2M}~ 
	(\mbox{\boldmath $p$}_j ~e^{-i \mbox{\boldmath $q$} \cdot
	\mbox{\boldmath $x$}_j}+ ~e^{-i 
	\mbox{\boldmath $q$} \cdot \mbox{\boldmath
	$x$}_j}~\mbox{\boldmath $p$}_j) \right. \nonumber \\ 
	& & \left. + \,\frac{1}{2}(1+\tau)_j ~i \left[ \frac{F_{1p}(q_\mu^2)
	+ \kappa_p 
	 F_{2p}(q_\mu^2)}{2M}\right]
	 \mbox{\boldmath $\sigma$}_j \times \mbox{\boldmath $q$}
	 ~e^{-i \mbox{\boldmath $q$} \cdot \mbox{\boldmath $x$}_j} \right]
	|\Psi_A \rangle|u_i\rangle \,.
\end{eqnarray}
Here the subscript $p$ refers to the proton. The transition matrix 
can be written in the convenient form as 
\begin{equation}
	{\cal M} = -\bigg[\langle u_f|u_i\rangle Q - \langle
	u_f|\mbox{\boldmath 
	$\alpha$}|u_i\rangle  \cdot {\bf J}\bigg]\,,
\end{equation}
where
\begin{eqnarray}
	Q & = & F_{\rm ch}^p(1+q_\mu^2/8M^2)\left\langle f\left|\sum_{j=1}^A
	e^{-i \mbox{\boldmath $q$} \cdot
	\mbox{\boldmath
	$x$}_j}\,\frac{1}{2}(1+\tau_3)_j\right|\Psi_A\right\rangle , 
	\nonumber\\ 
	{\bf J} & = &{\bf J}^{\rm el}+{\bf J}^{\rm mag},\nonumber \\
	{\bf J}^{\rm el} & = & \left\langle f\left|\sum_{j=1}^A(F_{\rm
	ch}^p/2M) 
	(\mbox{\boldmath $p$}_j e^{-i \mbox{\boldmath $q$} \cdot
	\mbox{\boldmath $x$}_j}+e^{-i 
	\mbox{\boldmath $q$} \cdot \mbox{\boldmath
	$x$}_j}\mbox{\boldmath $p$}_j)		 
	\,\frac{1}{2}(1+\tau_3)_j\right|\Psi_A\right\rangle ,\nonumber\\
	{\bf J}^{\rm mag} & = & (i/2M)F_{\rm mag}^p\left\langle
	f\left|\sum_{j=1}^A 
	 e^{-i \mbox{\boldmath $q$} \cdot \mbox{\boldmath
	$x$}_j}\mbox{\boldmath 
	$\sigma$}\times\mbox{\boldmath $q$} 
	\,\frac{1}{2}(1+\tau_3)_j\right|\Psi_A\right\rangle\,. 
\label{eq:jel}
\end{eqnarray}
In these equations, $\tau_3$ is the nucleon isospin operator, and
$F_{\rm ch}^p$ and $F_{\rm mag}^p$ are the proton charge and magnetic form
factors defined by
\begin{eqnarray}
	F_{\rm ch}^p & = & F_{1p} + (q_\mu^2/4M^2)\kappa_p F_{2p} \\
	F_{\rm mag}^p & = & F_{1p} + \kappa_p F_{2p}\,.
\end{eqnarray}
The analytical fit to the proton form factors $F_{1p}$ and $F_{2p}$,
given by Janssens {\em et al}. \cite{Janssens} is used in the calculations.

Squaring the matrix element and summing and averaging over electron
 spins yields 
\begin{eqnarray}
	\frac{1}{2} \sum_{\rm electron\;\;spins}|{\cal M}|^2
      	& = &  [(4E_iE_f + q_\mu^2)QQ^*
	    - q_\mu^2{\bf J} \cdot {\bf J}^* \nonumber \\
      	&   & + 2({\mbox{\boldmath $k$}}_f\cdot{\bf
	J})({\mbox{\boldmath $k$}}_i\cdot{\bf J}^*) 
	    + 2({\mbox{\boldmath $k$}}_f\cdot{\bf
      	J}^*)({\mbox{\boldmath $k$}}_i\cdot{\bf J})\nonumber\\ 
      	&   & -2E_f\{({\mbox{\boldmath $k$}}_i\cdot{\bf J})Q^* +
      	({\mbox{\boldmath $k$}}_i\cdot{\bf J}^*)Q\} \nonumber \\
      	&   & -2E_f\{({\mbox{\boldmath $k$}}_f\cdot{\bf J})Q^* +
      	({\mbox{\boldmath $k$}}_f\cdot{\bf J}^*)Q\}\,.
\end{eqnarray}
Substituting this expansion into Eq.~(\ref{eq:tcross}) results in
\begin{eqnarray}
	\frac{d^3\sigma}{dE_f\,d\Omega_p\,d\Omega_e} & =
	&\frac{\sigma_{\rm M}}{(\hbar c)^3 (2\pi)^3}~
	\frac{|\mbox{\boldmath $p$}_p|E_p}{1 - \frac{\displaystyle
	E_p}{\displaystyle 
	E_{A-1}}~\frac{\mbox{\boldmath $p$}_p 
	\cdot \mbox{\boldmath $p$}_{A-1}}{|\mbox{\boldmath
	$p$}_p|^2}}\nonumber \\ 
	& & \left\{|Q|^2 - \frac{1}{2} \sec^2 \frac{\theta}{2}~(Q^*{\bf
	J}+{\bf J}^*Q) 
	\cdot(\hat{k}_i + \hat{k}_f) \right. \nonumber \\
	& & +\frac{1}{2} \sec^2 \frac{\theta}{2}~({\bf J}\cdot \hat{k}_i
	{\bf J}^* \cdot 	
	\hat{k}_f + {\bf J}\cdot \hat{k}_f {\bf J}^* \cdot \hat{k}_i)
	\nonumber \\ 
	& & + \left. |{\bf J}|^2 \tan^2 \frac{\theta}{2}\right\}\,.
	\label{eq:xsec}
\end{eqnarray}
The evaluation of the coincidence cross section is thus reduced to the
evaluation of the nuclear matrix elements $Q$ and ${\bf J}$, which depend on
the choice of the initial and final wave functions.

%\subsection{Scattering States}

The inclusion of the FSI presupposes that the full solution of the
scattering state be determined. Although this has been achieved in the
three--nucleon case, it has not yet been accomplished  
in the four--nucleon case beyond the break up threshold.
To circumvent this problem we employ the
model outlined in the introduction in which the FSI is taken into
account via an optical potential treatment of the relative motion of
the outgoing clusters.  In this particularly simplifying modification
the plane wave $\langle\mbox{\boldmath $q$}_\alpha|$ is replaced by
scattering states $^{(-)}\langle\mbox{\boldmath $q$}_\alpha|$
generated by a $p-^3$H potential obtained using the Marchenko inverse
scattering procedure \cite{AM63,chadsab,Alt}.  In other
words, instead of solving a four-body integral equation providing us
with the full scattering state $^{(-)}\langle \mbox{\boldmath
$q$}_\alpha;\psi_B|$, we use the approximation given by Eq.~(\ref{eq:scatt}).

In implementing this model, we found it was 
numerically more stable to solve the integral equations in momentum
space, in a similar fashion to the method employed by Fiedeldey {\em
et al}.\ \cite{Fiedeldey} in  the photodisintegration of $^4$He. In
this method, the scattering state and the channel state can be
projected onto each other by means of M{\o}ller operators
\cite{Sandhas}. The resulting equation 
is of the Lippmann-Schwinger-type and can be written schematically as
\begin{equation}
	|{\cal M}\rangle = |{\cal B}\rangle + VG_0|{\cal M}\rangle\,.
\label{eq:ls}
\end{equation}
$|{\cal B}\rangle$ is the plane--wave (Born) term,
which can be evaluated in either configuration
or momentum space using the wave functions described below. 

%\subsection{Bound States}

The three-- and four--nucleon bound state spatial wave functions are
obtained using the IDEA. In this method, the $A$-nucleon wave function
$\Psi(\mbox{\boldmath $x$})$ is written as a sum of subamplitudes
$\Psi(\mbox{\boldmath $x$}) = \sum_{i<j\leq A}\psi(\mbox{\boldmath
$r$}_{ij},r)$ obeying the Faddeev-type equation  
\begin{equation} 
	[T+\frac{A(A-1)}{2}V_0(r)]\psi(\mbox{\boldmath
	$r$}_{ij},r)=-[V(\mbox{\boldmath 
	$r$}_{ij})-V_0(r)] 
	\sum_{i<j\leq A}\psi(\mbox{\boldmath $r$}_{ij},r) 
\label{eq:idea1} 
\end{equation}
where $r$ is the hyperradius of the system, $r^2 =
\frac{2}{A}\sum\mbox{\boldmath $r$}^2_{ij}$ and
$V_0(r)$ is the hypercentral potential which is the first term of
the Potential Harmonic expansion of the interaction. 
The effects of the all important higher partial waves are
approximately included via the use of the hypercentral potential
(for more details of the IDEA see Refs.~\cite{IDEA1,Oehm1,Oehm}).

\section{Results}
\label{result}
\begin{figure}[htb]
\centerline{\epsfig{file=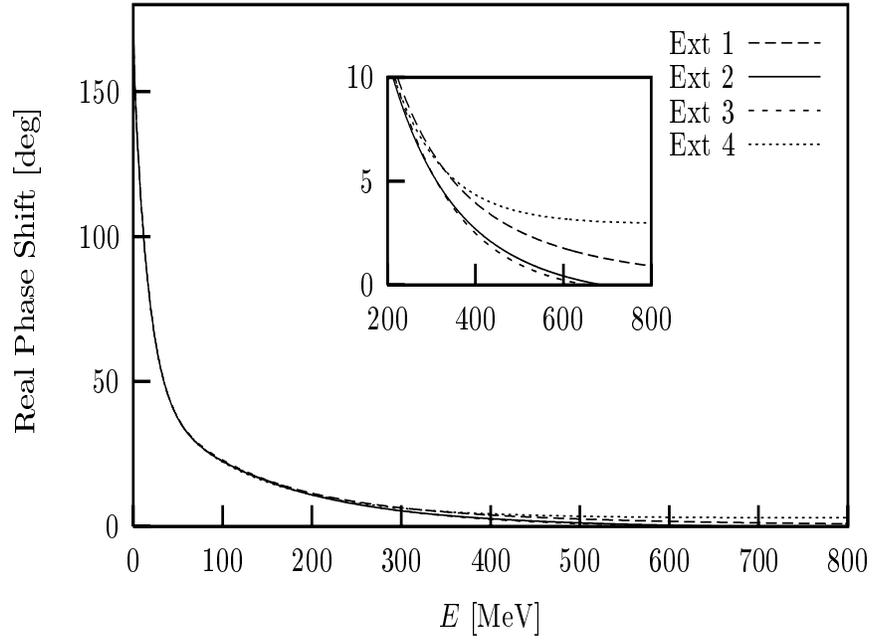,height=9cm,width=12cm}}
\caption{\label{del4}  Typical extrapolations of the high energy $p+^3$H
phase shifts used.}
\end{figure}

\begin{figure}[htb]
\centerline{\epsfig{file=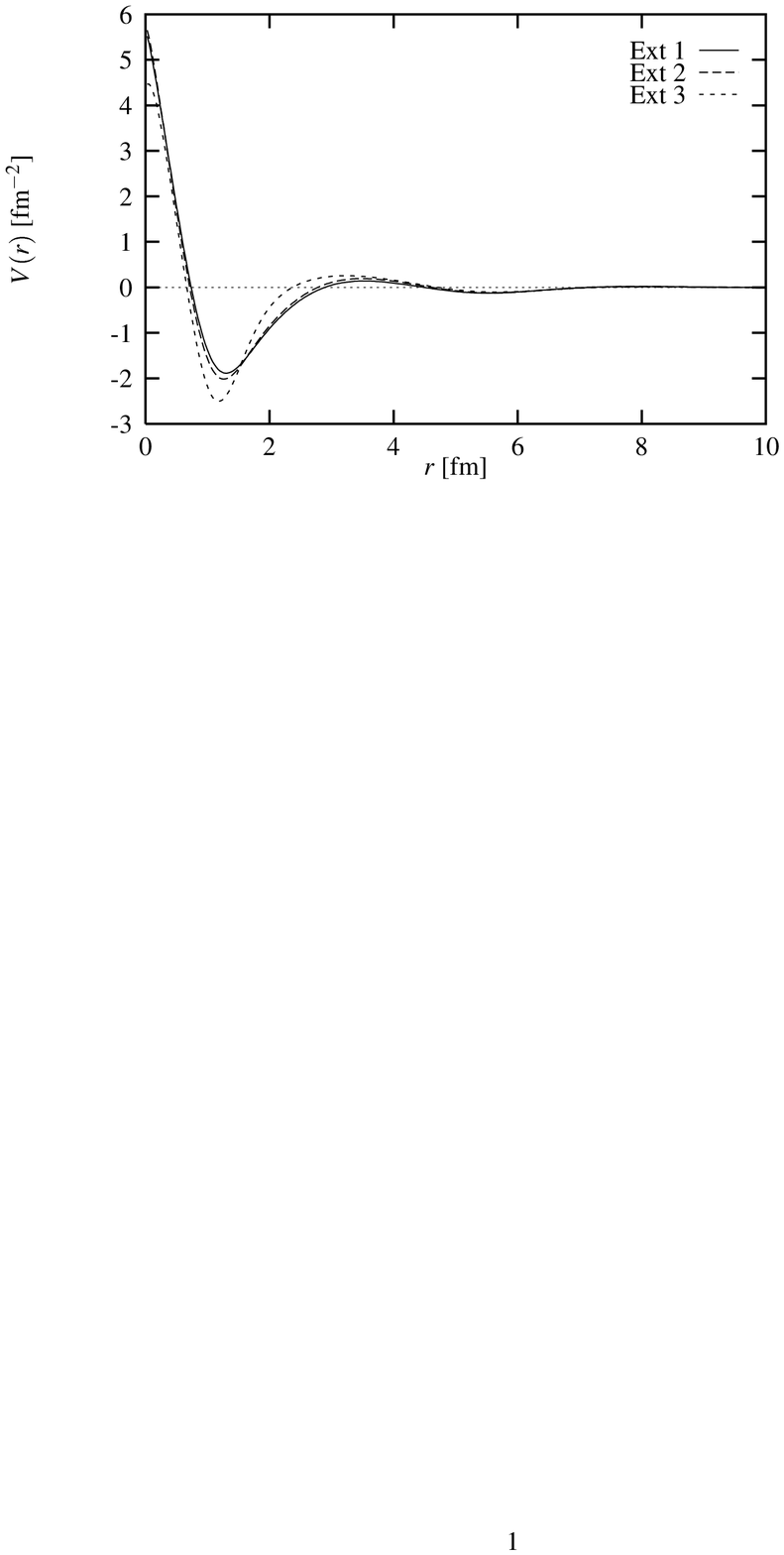,height=9cm,width=12cm}}
\caption{\label{vpot4r}  Real part of the Marchenko potentials
corresponding to the phase shifts  given in Fig.~1.}
\end{figure}

Reliable experimental phase shifts required in the construction of the
effective potential for the 3+1 fragmentation of
$^4$He are not available. Furthermore, the energy region in which they
are available has little or no overlap with the region
needed for our calculations. On the theoretical front, phase 
shifts beyond the three--body break up have not yet been 
obtained. Therefore, in implementing our model we 
employed phase shifts obtained from the optical potential 
used by Schiavilla \cite{Schia4,Oers}.    
Since the phase shifts produced by this potential become negative
for energies greater than 200\,MeV, we had to extrapolate them 
to higher energies. In Fig.~\ref{del4} four such extrapolations of 
the real phase shifts to higher energies are plotted. 
The corresponding real parts of the Marchenko potentials that describe
the $p+^3$H interaction are plotted in Fig.~\ref{vpot4r}.
The local potentials produced by the Marchenko
procedure are complex and unique for each set of data. They are phase
equivalent up to approximately 300\,MeV, producing phase shifts with
slightly different high energy behavior. The extrapolation Ext. 2,
produced an overlap integral that was, for  all practical purposes, zero.\par

\begin{figure}[htb]
\centerline{\epsfig{file=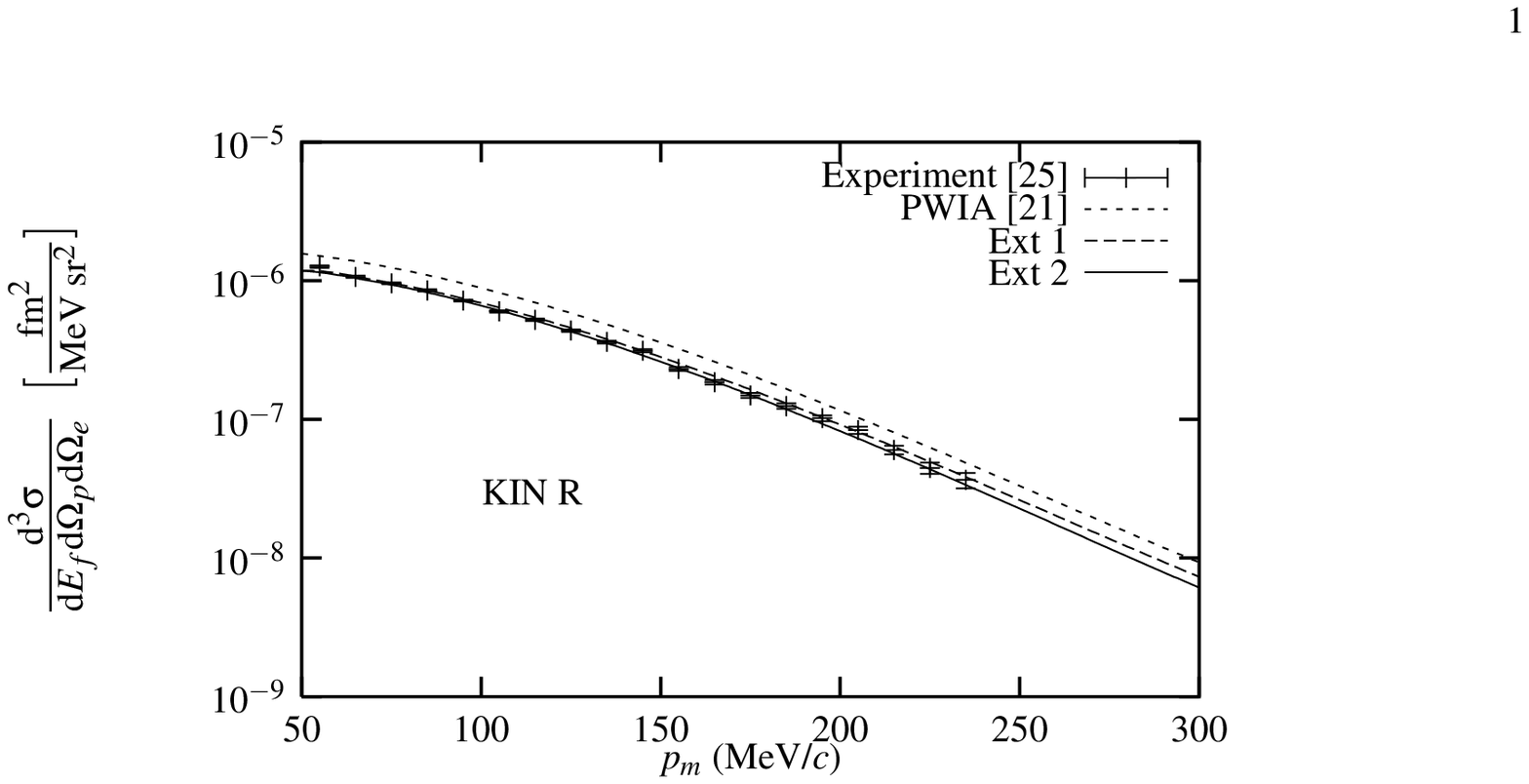,height=9cm,width=12cm}}
\caption{\label{kinR}  The two-fragment electrodisintegration cross
section of $^4$He as a function of the missing momentum for kinematics
R of Table~1.}
\end{figure}

The various kinematics used are given in Table \ref{kin4}.  In 
Fig.~\ref{kinR} we have plotted  our cross section results for the
kinematics R together with the experimental data of
Ref.~\cite{Koos}. These results are very close to those obtained
by Laget \cite{Laget4}. It is obvious that the cross section is only
weakly dependent on the nucleon--nucleus interaction, all
extrapolations yielding virtually identical results. \par

\begin{figure}[htb]
\centerline{\epsfig{file=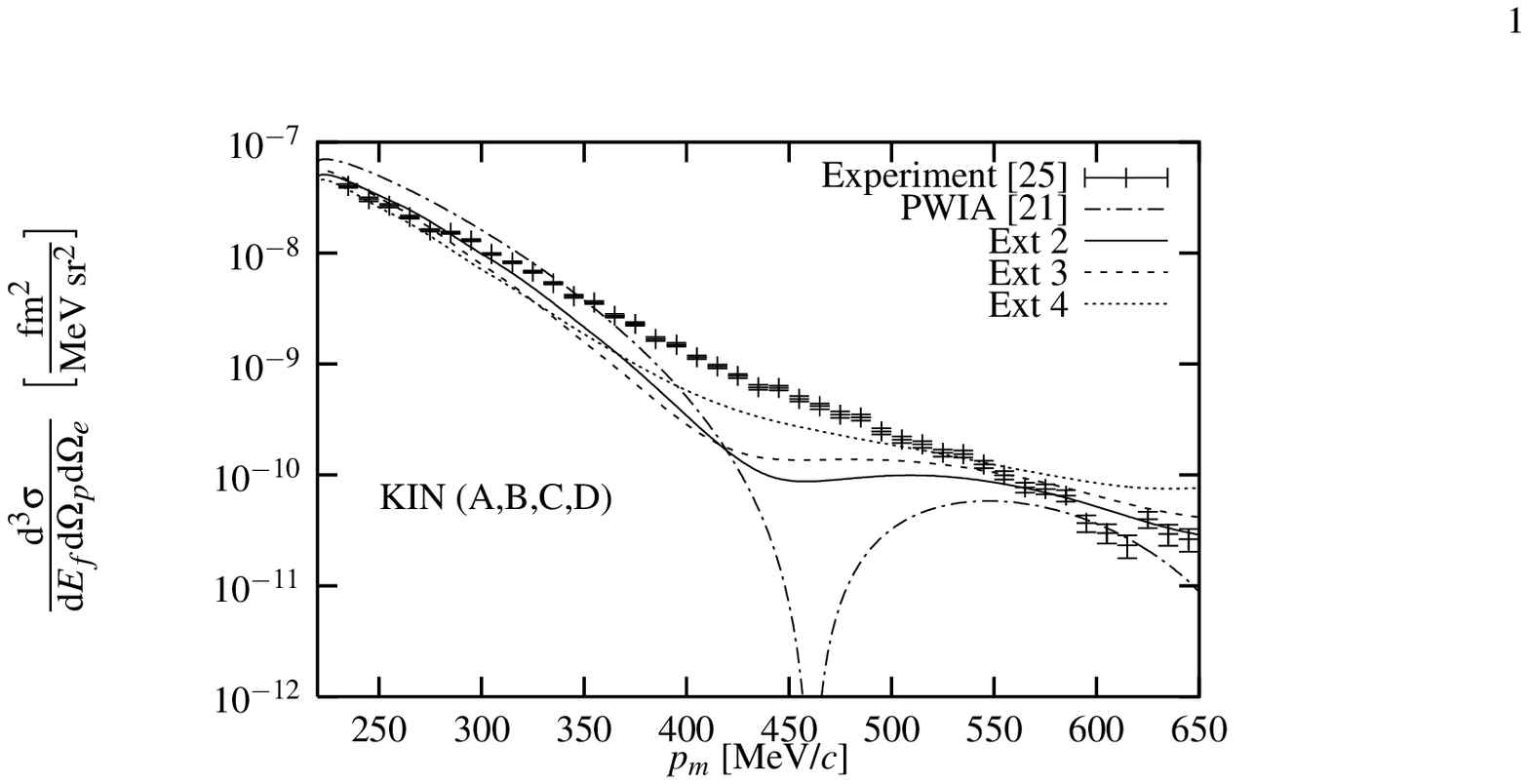,height=9cm,width=12cm}}
\caption{\label{kinABCD}  The two-fragment electrodisintegration cross
section of $^4$He as a function of the missing momentum for kinematics
(A,B,C,D) of Table~1.} 
\end{figure}

\begin{figure}[htb]
\centerline{\epsfig{file=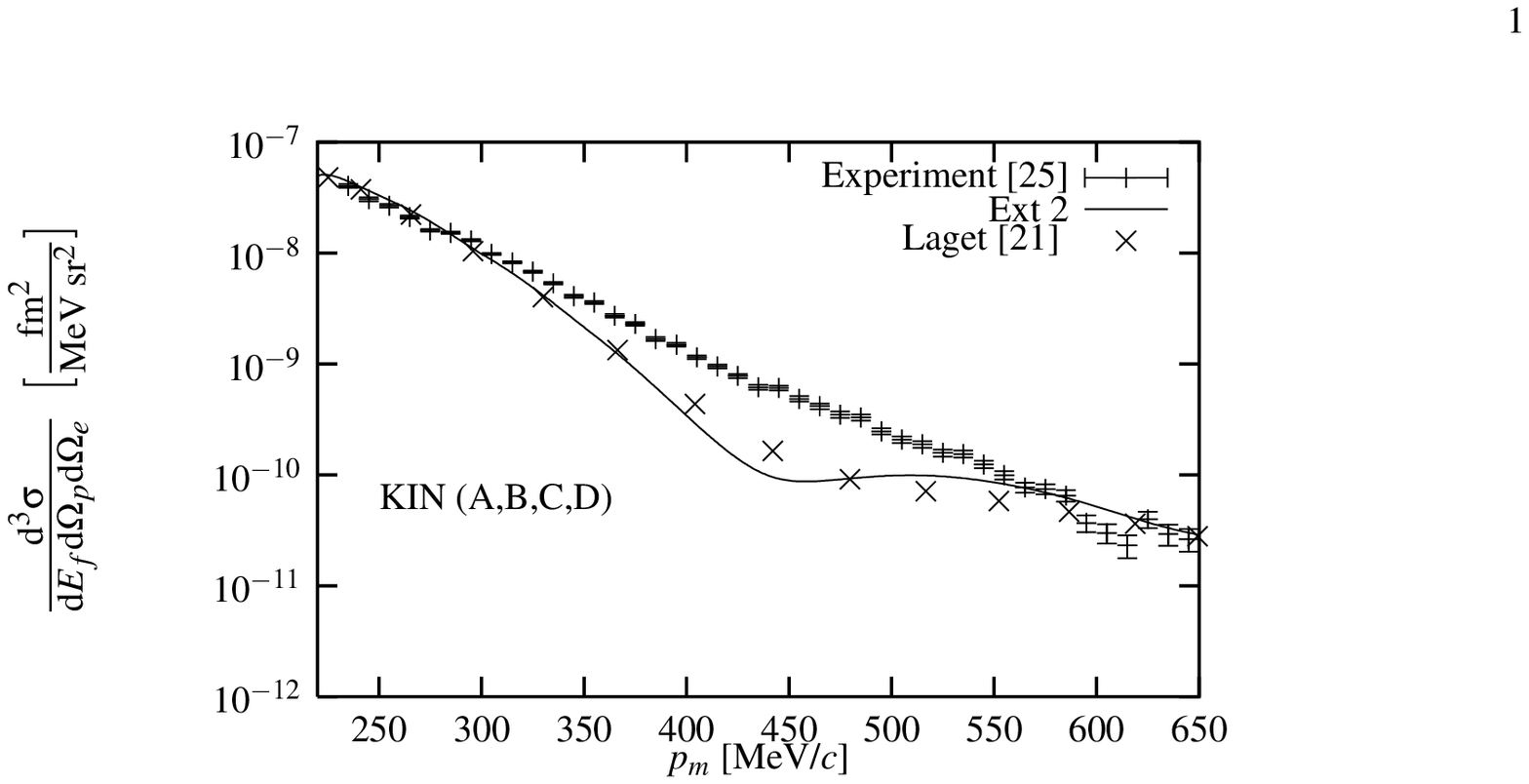,height=9cm,width=12cm}}
\caption{\label{koosabc_comp}  A comparison of our two-fragment
electrodisintegration cross section of $^4$He with that of Laget as a
function of the missing momentum for kinematics (A,B,C,D).}
\end{figure}

\begin{figure}[htb]
\centerline{\epsfig{file=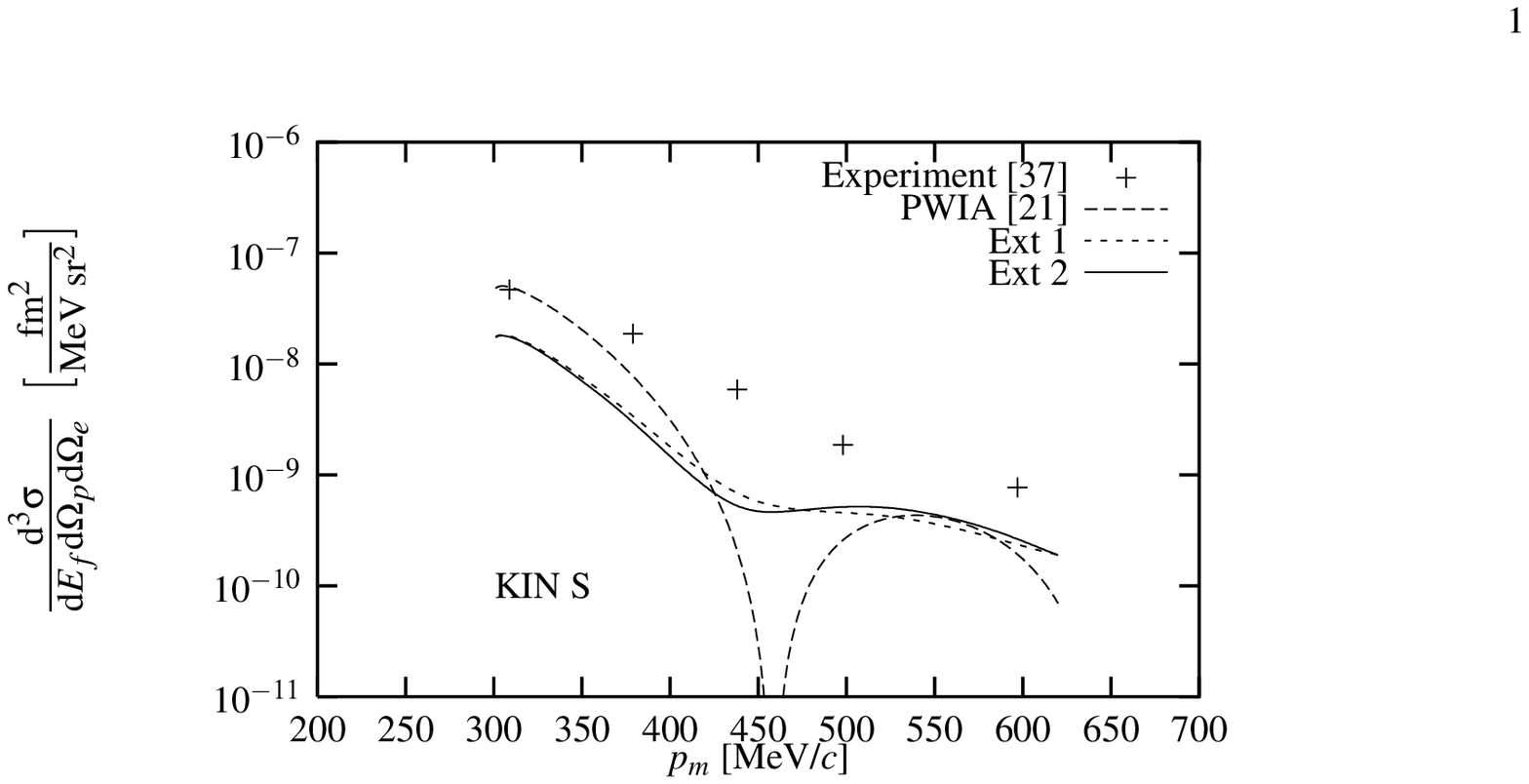,height=9cm,width=12cm}}
\caption{\label{kinS}  The two-fragment electrodisintegration cross
section of $^4$He as a function of the missing momentum for kinematics
S of Table~1. }
\end{figure}

The effect of the high--energy behavior of the phase shifts on the
cross section and the importance of the orthogonality condition given
by Eq.~(\ref{eq:overlap}), is more clearly demonstrated in the plots
of the cross sections for the kinematics (A,B,C,D) and kinematics S,
shown in Figs.~\ref{kinABCD} to \ref{kinS}. The PWIA prediction for
kinematics (A,B,C,D) shows a minimum, in fact a zero, in agreement 
with the calculations of Laget and Schiavilla. At missing momenta 
below 300\,MeV/$c$, all calculations show good agreement with the
data, but in the region of the dip  the calculations underestimate the data
considerably. Beyond missing momenta of $\sim$ 550\,MeV/$c$ the
agreement of the calculations with the data is again fair. The best
result is the one obtained for Ext. 2 which fulfills the
orthogonality condition (\ref{eq:overlap}). In
Fig.~\ref{koosabc_comp}, we compare the cross section produced using
this extrapolation with the results of Laget \cite{Laget4};  it is
seen that the agreement is good. \par

The PWIA cross section calculated for kinematics S, shown in
Fig.~\ref{kinS}, also exhibits a dip at $\sim$450\,MeV/$c$. 
Including the FSI causes a partial filling of 
this dip. The position of the  minimum and the value
of the cross section in this region once again depends strongly on the
input nucleon--nucleus and the NN potentials.

\section{Conclusions}
\label{concl}
The results described above clearly indicate a sensitivity of the cross
section in certain kinematical regions to the input nucleon--nucleus
interaction  and to the input NN potential. In our model, 
the various nucleon-nucleus (3+1) interactions
employed for this investigation were phase equivalent up to
approximately 300\,MeV. They all reproduced the correct binding energy
for the four--nucleon bound state. However, beyond 300\,MeV,
the potentials produced phase shifts that differed slightly. The
corresponding nucleon--nucleus potentials provided scattering states
that were not always orthogonal to the initial bound state. This in turn  is
translated into large differences in the cross sections in certain
kinematical regions.  This result was not surprising, as already in
1970, Fiedeldey \cite{Fied} investigated the dependence of the triton
binding energy on the high-energy part of the phase shift. He concluded that
arbitrary variations of the NN phase shift at high energies ($E_{\rm
lab}>300$\,MeV) can produce large differences in the triton binding
energy.  This sensitivity is also manifested in the
electro-disintegration cross sections which implies that we need a much
clearer idea of what constitutes a physically acceptable extension of
the phase shift to higher energies. This sounds a warning that in
model calculations care must be taken when using effective
interactions, the most important constraint being the condition that
the initial bound state and the final scattering state be orthogonal.\par

The dip  around 450\,MeV/$c$, which is present in
all model calculations, is something of a surprise. At missing momenta 
less than 300\,MeV/$c$ all calculations show a good agreement with 
the experimental data, {\em i.e.}, the PWIA performs reasonably well 
in a region where the FSI could be expected to be more important.  
At missing momenta beyond the dip area, where the MEC become important,
the agreement with the data is fair. Thus the zero in the 
PWIA cross-section is not necessarily a manifestation of strong
FSI effects or MEC. It can simply  be attributed to the vanishing 
Born term which contains only genuine $3+1$ components. The Born term
should, however, be treated in a more rigorous way 
using the AGS formalism which allows for the effects coming from the 2+2
channel. The inclusion of this channel should result in an improvement
of the the bound state wave functions and of the Born term. This
in turn is expected to  remove the apparent zero in the PWIA
cross section. Investigations in this regard are currently under way.

%%%%%%%%%%%%%%%%%%%%%%%%%%%%%%%%%%%%%%%%%%%%%%%%%%%%%%%%%%%
%%%%%%%%%%%%%%%%      TABLE    %%%%%%%%%%%%%%%%%%%%%%%%%%%%
%%%%%%%%%%%%%%%%%%%%%%%%%%%%%%%%%%%%%%%%%%%%%%%%%%%%%%%%%%%
\begin{table}[htb]
\caption{\label{kin4} Kinematics R and (A,B,C,D) from
Ref.~[25], and kinematics S from Ref.~[37].} 
\begin{center}
\begin{tabular}{|c||c|c|c|c|} 
\hline
       &  $E_i$ [MeV]    &  $E_f$ [MeV]     & $\theta$  & $q$ [MeV/$c$]
                                                 \\ \hline
    R  &  524.9   &  423.9    & 49.60$^o$  & 408\\
(A,B,C,D)&  525.0   &  310.0    & 49.60$^o$  & 401\\ \hline
    S  &  560.0   &  360.0    & 25.00$^o$  & 278\\ \hline
\end{tabular}
\end{center}
\end{table}

%%%%%%%%%%%%%%%%%%%%%%%%%%%%%%%%%%%%%%%%%%%%%%%%%%%%%%%%%%%
%%%%%%%%%%%%%%%%   FIGURES     %%%%%%%%%%%%%%%%%%%%%%%%%%%%
%%%%%%%%%%%%%%%%%%%%%%%%%%%%%%%%%%%%%%%%%%%%%%%%%%%%%%%%%%%

\end{document}